\documentclass[aps,twocolumn,longbibliography]{revtex4-1}
\pdfoutput=1

\usepackage{amsmath,amssymb,mathtools,bm,color,graphicx,empheq}

\newcommand{\imag}{\operatorname{Im}}
\newcommand{\sinc}{\operatorname{sinc}}

\begin{document}

\title{Quantum backflow in a ring}

\author{Arseni Goussev}

\affiliation{School of Mathematics and Physics, University of Portsmouth, Portsmouth PO1 3HF, United Kingdom}

\date{\today}

\begin{abstract}
	Free motion of a quantum particle with the wave function entirely comprised of plane waves with non-negative momenta may be accompanied by negative probability current, an effect called quantum backflow. The effect is weak and fragile, and has not yet been observed experimentally. Here we show that quantum backflow becomes significantly more pronounced and more amenable to experimental observation if, instead of letting the particle move along a straight line, one forces it to move in a circular ring.
\end{abstract}

\maketitle

\section{Introduction}

The probability density of a quantum particle may flow in the direction opposite to that of the particle's momentum \cite{All69time-c, Kij74time}, an effect called quantum backflow (QB) \cite{BM94Probability} \footnote{See Ref.~\cite{YH13introduction} for a short introduction to the quantum backflow effect.}. The effect is inconceivable from the viewpoint of classical physics, and in this respect can be paralleled with other genuinely quantum phenomena such as tunneling or Schr\"odinger's cat states. Unlike the latter, however, QB is relatively unexplored and yet to be observed experimentally.

The QB effect can be formulated as follows. Consider a nonrelativistic particle moving freely along a straight line, the $x$-axis. Let the particle's wave function be comprised only of plane waves $\frac{1}{\sqrt{2 \pi}} e^{i k x}$ with nonnegative momenta, $\hbar k \ge 0$, so that at time $t$ the wave function reads
\begin{equation}
	\psi(x,t) = \int_0^{\infty} dk \, \phi(k) e^{-i \hbar k^2 t / 2 \mu} \frac{e^{i k x}}{\sqrt{2 \pi}} \,,
\end{equation}
where $\mu$ is the particle's mass, and $\phi(k)$ is a complex-valued function normalized according to $\int_0^{\infty} dk \, |\phi(k)|^2 = 1 = \int_{-\infty}^{+\infty} dx \, |\psi(x,t)|^2$. The associated probability current $j_{\psi}(x,t)$ is given by
\begin{equation}
	j_{\psi} = \frac{\hbar}{\mu} \imag \left\{ \psi^* \frac{\partial \psi}{\partial x} \right\} \,.
\end{equation}
The QB effect consists in the fact that $j_{\psi}$ can be negative, for some $x$ and $t$, in spite of the particle's momentum being nonnegative with certainty. In other words, even though the momentum of a particle is pointing ``to the right'', the probability density can (locally in space and time) flow ``to the left''; this is clearly impossible in the classical world.

One of the most surprising features of QB is that the effect has a
nontrivial dimensionless scale associated with it: The probability current through a given point, say $x = 0$, integrated over an arbitrary time window, $-T/2 < t < T/2$, has a (finite) greatest lower bound. More precisely \cite{BM94Probability, EFV05Quantum, PGKW06new},
\begin{equation}
	\inf_{\psi} \int_{-T/2}^{T/2} dt \, j_{\psi}(0,t) = -c_{\text{line}} \,,
\end{equation}
where
\begin{equation}
	c_{\text{line}} \simeq 0.0384517
\label{BM_bound}
\end{equation}
is the so-called Bracken-Melloy bound. Finding the exact value of $c_{\text{line}}$ remains an open challenge. It is interesting to note that $c_{\text{line}}$ is independent of the time window $T$, the particle's mass $\mu$, or Planck's constant $\hbar$.

Many questions related to QB have been addressed in the literature. These include QB against a constant force~\cite{MB98velocity}, the pertinence of QB to the arrival-time problem \cite{MPL99Arrival, ML00Arrival, DEHM02Measurement, HBLO19Quasiprobability}, position dependence of the backflow current~\cite{EFV05Quantum, Ber10Quantum, BCL17Quantum}, probability backflow in relativistic quantum systems~\cite{MB98Probability, SC18Quantum, ALS19Relativistic}, QB in escape problems~\cite{Gou19Equivalence, DT19Decay}, and QB in many-particle systems~\cite{Bar20Quantum}. Recently, the problem of QB has been generalized to states with position-momentum correlations \cite{Gou20Probability}.

As of today, QB has not been experimentally observed in any true quantum system \footnote{An experimental scheme for observing QB in Bose-Einstein condensates was proposed in Ref.~\cite{PTMM13Detecting}. Also, there has been a recent experimental realization of an optical analogue of the QB effect \cite{EZB20Observation}.}. One of the difficulties hindering experimental observation of QB is a relatively small value of $c_{\text{line}}$ \cite{[This difficulty is aggravated by the fragility of the QB effect in the presence of thermal noise. See: ]AGP16Quantum}. As pointed out in Ref.~\cite{BM94Probability}, one natural strategy for detecting QB would be to work with an electrically charge particle, for which a measurement of the probability current is equivalent to that of the electric current. If the charge of the particle is $q$ (for concreteness taken to be positive) and the current measurement time is $T$, then the magnitude of the detected backflow electric current approximately equals $-\frac{q}{T} \int_{-T/2}^{T/2} dt \, j_{\psi}$, which cannot exceed $c_{\text{line}} q / T$. The fact that $c_{\text{line}}$ is less than $4\%$ hampers the direct detection of the backflow current.

Another obstacle to observing QB experimentally is the difficulty of preparing a state with an appreciable backflow current. Theoretical considerations show \cite{YHHW12Analytical} that states whose integrated backflow current is close to the Bracken-Melloy bound, $c_{\text{line}}$, are characterized by infinite position width and infinite mean energy, and therefore are not realizable in a laboratory setting. The value of the integrated backflow current seems to become significantly smaller than $c_{\text{line}}$ if one restricts their attention to the states with a finite position width and/or finite mean energy (see Refs.~\cite{BM94Probability, HGL+13Quantum} for some examples), although no systematic study of this question has yet been carried out.

In this paper we show that the QB effect becomes much more pronounced and more amenable to experimental investigation when considered for a quantum particle moving in a circular ring. In particular, we show that, for the particle-in-a-ring system, the integrated backflow current can be over three times larger than the Bracken-Melloy bound, and that the corresponding backflow maximizing state has finite energy (and, by construction, finite spatial extent). Some space-related aspects of QB in systems with rotational motion, such as an electron in a constant magnetic field, have been addressed in Ref.~\cite{Str12Large} and very recently in Ref.~\cite{PPR20Angular}. Here however we are interested in the time-dependence of QB, and more specifically look for states maximizing the integrated backflow current. 

\section{Particle in a ring}

We consider a nonrelativistic particle of mass $\mu$ constrained to move in a circular ring of radius $R$. The ring lies in the $xy$-plane of a Cartesian coordinate frame and is centered around the origin. The unit vectors along the $x$-, $y$- and $z$-axes are denoted by $\bm{e}_x$, $\bm{e}_y$, and $\bm{e}_z$, respectively. The triplet $(\bm{e}_x, \bm{e}_y, \bm{e}_z)$ is right-handed. We further assume that the particle has an electric charge $q$, for concreteness taken to be positive, and that there is a constant spatially-uniform magnetic field $\bm{B}$ pointing along the $z$-axis, i.e., $\bm{B} = B \bm{e}_z$.

The particle is described by a time-dependent wave function $\Psi(\theta, t)$, where $\theta$ is the polar angle between $\bm{e}_x$ and the position radius vector of the particle. The wave function is periodic, $\Psi(\theta + 2 \pi, t) = \Psi(\theta, t)$, and satisfies the Schr\"odinger equation $i \hbar \frac{\partial \Psi}{\partial t} = H \Psi$ with the Hamiltonian \cite{Sca02Classical, Vug04carged}
\begin{equation}
	H = \frac{\hbar^2}{2 \mu R^2} (\ell_z - \beta)^2 \,,
\label{Hamiltonian}
\end{equation}
where $\ell_z = -i \frac{\partial}{\partial \theta}$, so that $\hbar \ell_z$ is the projection of the canonical angular momentum on $\bm{e}_z$. Here,
\begin{equation}
	\beta = \frac{q R^2 B}{2 \hbar c}
\label{beta}
\end{equation}
is the dimensionless magnetic flux through the ring, with $c$ denoting the speed of light. (A constant term $\frac{\hbar^2}{8 \mu R^2}$ has to be added to the Hamiltonian, Eq.~\eqref{Hamiltonian}, if the latter is derived using the Dirac method \cite{Sca02Classical}. This term however plays no role in the context of the present work.) The wave function is normalized to unity,
\begin{equation}
	\int_0^{2 \pi} d\theta \, |\Psi(\theta,t)|^2 = 1 \,.
\label{norm-1}
\end{equation}
The probability density, $|\Psi(\theta,t)|^2$, satisfies the continuity equation:
\begin{equation}
	\frac{\partial |\Psi|^2}{\partial t} + \frac{\partial J_{\Psi}}{\partial \theta} = 0 \,,
\end{equation}
where $J_{\Psi}(\theta, t)$ is the probability current, defined as
\begin{equation}
	J_{\Psi} = \frac{\hbar}{\mu R^2} \imag \left\{ \Psi^* \frac{\partial \Psi}{\partial \theta} \right\} - \frac{\hbar \beta}{\mu R^2} |\Psi|^2 \,.
\label{prob_current}
\end{equation}
The Schr\"odinger equation, probability density, and probability current are invariant under the gauge transformation $\beta \to \beta + \frac{\partial \chi}{\partial \theta}$ and $\Psi \to e^{i \chi} \Psi$, where $\chi(\theta)$ is an arbitrary real function. Eigenstates $\psi_m$ and eigenenergies $E_m$ of the Hamiltonian satisfy $H \psi_m = E_m \psi_m$ and are given by
\begin{equation}
	  \psi_m(\theta) = \frac{e^{i m \theta}}{\sqrt{2 \pi}} \,, \qquad E_m = \frac{\hbar^2}{2 \mu R^2} (m - \beta)^2  \qquad (m \in \mathbb{Z}) \,.
\label{eig}
\end{equation}
The set of eigenstates is orthonormal and complete.

All Hamiltonian eigenstates $\psi_m$ with $m \ge \lceil \beta \rceil$, where  $\lceil \cdot \rceil$ is the ceiling function, have nonnegative (gauge-invariant) kinetic angular momentum and probability current. Indeed, $\psi_m$ is an eigenstate of the kinetic angular momentum operator $\hbar (\ell_z - \beta)$ with the eigenvalue $\hbar (m - \beta) \ge 0$, and the probability current corresponding to $\psi_m$ is $J_{\psi_m} = \frac{\hbar}{2 \pi \mu R^2} (m - \beta) \ge 0$. Now, in the spirit of the original QB problem, we consider states $\Psi(\theta,t)$ comprised of the Hamiltonian eigenstates with nonnegative kinetic angular momentum:
\begin{equation}
	\Psi(\theta,t) = \sum_{m = \lceil \beta \rceil}^{\infty} c_m \psi_m(\theta) e^{-i E_m t / \hbar} \,,
\label{wave_function}
\end{equation}
where, in view of Eq.~\eqref{norm-1}, complex amplitudes $c_m$ satisfy the normalization condition
\begin{equation}
	\sum_{m = \lceil \beta \rceil}^{\infty} |c_m|^2 = 1 \,.
\label{norm-2}
\end{equation}
The corresponding probability current is obtained by substituting Eq.~\eqref{wave_function} into Eq.~\eqref{prob_current}, and reads
\begin{align}
	J_{\Psi}(\theta,t)
	&= \frac{\hbar}{2 \mu R^2} \sum_{m, n = \lceil \beta \rceil}^{\infty} (m + n - 2 \beta) \nonumber \\
	&\qquad \times c^*_m c_n \psi^*_m(\theta) \psi_n(\theta) e^{i (E_m - E_n) t / \hbar} \,.
\label{prob_current-2}
\end{align}

We now focus on the probability current through a fixed point on the ring, say $\theta = 0$, integrated over a time window  $-T/2 < t < T/2$, and define the dimensionless quantity
\begin{equation}
	P_{\Psi} = \int_{-T/2}^{T/2} dt \, J_{\Psi}(0,t) \,.
\label{P_def}
\end{equation}
Substituting Eqs.~\eqref{prob_current-2} and \eqref{eig} into Eq.~\eqref{P_def}, and evaluating the time integral, we find
\begin{equation}
	P_{\Psi} = \sum_{m,n = \lceil \beta \rceil}^{\infty} c^*_m K_{mn} c_n
\label{P_general}
\end{equation}
with
\begin{equation}
	K_{mn} = \frac{\alpha}{\pi} (m + n - 2 \beta) \sinc \big[ \alpha (m + n - 2 \beta) (m - n) \big] \,.
\label{K_def}
\end{equation}
Here,
\begin{equation}
	\alpha = \frac{\hbar T}{4 \mu R^2}
\end{equation}
is a (positive) dimensionless parameter, and the sinc function is defined as $\sinc z = \frac{\sin z}{z}$ if $z \not= 0$ and $\sinc 0 = 1$.

Our aim is to investigate the integrated probability current, Eq.~\eqref{P_general}, in view of the normalization condition, Eq.~\eqref{norm-2}. Since $P_{\Psi}$ is invariant with respect to the transformation $\beta \to \beta + 1$ and $c_m \to c_{m-1}$, $m \in \mathbb{Z}$, it is sufficient to only consider the parametric interval
\begin{equation}
	-1 < \beta \le 0 \,.
\label{beta_regime}
\end{equation}
On this interval $\lceil \beta \rceil = 0$, and so Eqs.~\eqref{P_general} and \eqref{norm-2} take the form
\begin{equation}
	P_{\Psi} = \sum_{m,n = 0}^{\infty} c^*_m K_{mn} c_n
\label{P_general-2}
\end{equation}
and
\begin{equation}
	\sum_{m = 0}^{\infty} |c_m|^2 = 1 \,,
\label{norm-3}
\end{equation}
respectively. Hereafter, we rely on Eqs.~\eqref{beta_regime}--\eqref{norm-3}.

It is worth nothing that $P_{\Psi}$ is unbounded from above. This is readily established by taking $c_m = \delta_{m m_1}$, with $m_1 \ge 0$, for which $P_{\Psi} = 2 \alpha (m_1 - \beta) / \pi$, and observing that $P_{\Psi} \to \infty$ as $m_1 \to \infty$. However, the nontrivial questions are {\it whether $P_{\Psi}$ can be negative}, and {\it whether $\inf P_{\Psi}$ is finite}.

We begin our study by considering an example scenario in which $\Psi$ is comprised of only two eigenstates, $\psi_{m_1}$ and $\psi_{m_2}$ with $0 \le m_1 < m_2$. Thus, we take
\begin{equation}
	c_m = \left\{
	\begin{array}{ll}
		\cos \frac{\varphi}{2} \quad &\text{if} \quad m = m_1 \ge 0 \\[0.1cm]
		e^{i \gamma} \sin \frac{\varphi}{2} \quad &\text{if} \quad m = m_2 > m_1 \\[0.1cm]
		0 &\text{otherwise}
	\end{array} \right.
\label{c_m_example}
\end{equation}
with $0 \le \varphi \le \pi$ and $0 \le \gamma < 2 \pi$. This parametrization ensures Eq.~\eqref{norm-3} is fulfilled. Substituting Eq.~\eqref{c_m_example} into Eq.~\eqref{P_general-2}, we obtain
\begin{equation}
	P_{\Psi} = \frac{\alpha}{\pi} \big[ A - B \cos \varphi + A \sinc(\alpha A B) \cos \gamma \sin \varphi \big] \,,
\end{equation}
where
\begin{equation}
	A = m_1 + m_2 - 2 \beta \,, \qquad B = m_2 - m_1 \,.
\end{equation}
We now look for the minimum of $P_{\Psi}$ with respect to $\varphi$ and $\gamma$ (for fixed values of $\alpha$, $\beta$, $m_1$, and $m_2$), i.e.
\begin{equation}
	\mathcal{P}^{(m_1, m_2)}(\alpha, \beta) = \min_{\varphi, \gamma} P_{\Psi} \,.
\end{equation}
A straightforward calculation yields
\begin{equation}
	\mathcal{P}^{(m_1, m_2)} = \frac{\alpha}{\pi} \left( A - \sqrt{B^2 + A^2 \sinc^2(\alpha A B)} \right) \,.
\label{P^(m1,m2)}
\end{equation}

\begin{figure}[h]
	\centering
	\includegraphics[width=0.45\textwidth]{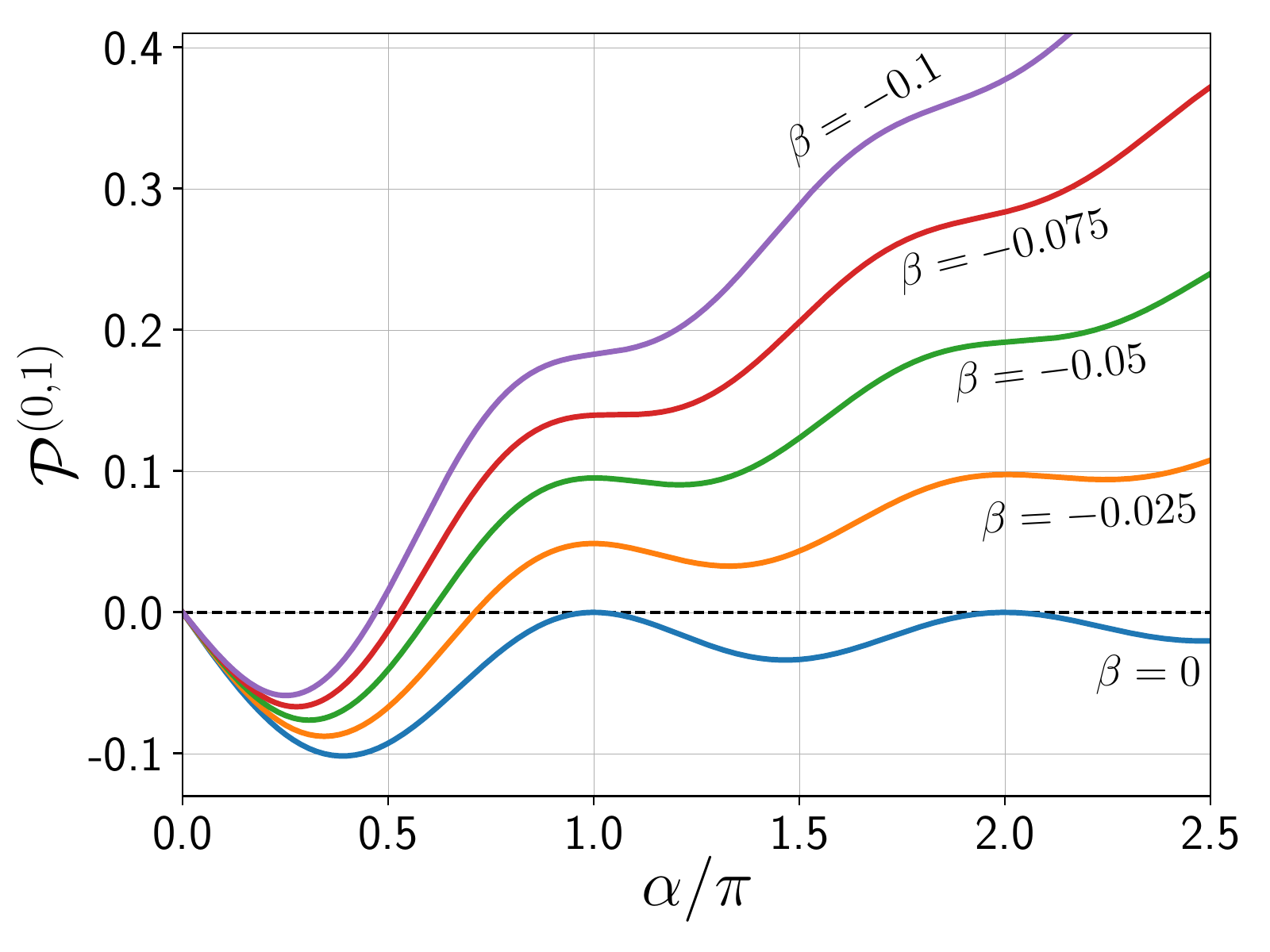}
	\caption{Minimum of the time-integrated probability current, Eq.~\eqref{P^(m1,m2)}, for $m_1 = 0$ and $m_2 = 1$, as a function of $\alpha$ for five different values of $\beta$.}
	\label{fig1}
\end{figure}

Figure~\ref{fig1} shows the dependence of $\mathcal{P}^{(0,1)}$ (corresponding to $m_1 = 0$ and $m_2 = 1$) on $\alpha$ for five different values of $\beta$. There are two main messages conveyed by this figure. First, it confirms that the integrated probability current can indeed be negative. Second, it shows that already for some very simple states (such as a superposition of $\psi_0$ and $\psi_1$) the magnitude of the integrated negative probability current can significantly exceed the Bracken-Melloy bound, $c_{\text{line}}$. In fact, numerical evaluation shows that $\min_{\alpha, \beta} \mathcal{P}^{(0,1)} \simeq -0.101727 \simeq -2.6 \times c_{\text{line}}$.

Consideration of cases other than $(m_1, m_2) = (0, 1)$ does not reveal a more pronounced backflow. It is easy to verify that for any $(m_1, m_2)$, such that $0 \le m_1 < m_2$,
\begin{align}
	&\mathcal{P}^{(m_1,m_2)}(\alpha, \beta) \nonumber \\ &\quad = \frac{1}{m_2 - m_1} \mathcal{P}^{(0,1)} \left( \alpha (m_2 - m_1)^2, \frac{\beta - m_1}{m_2 - m_1} \right) \,.
\end{align}
This scaling relation, in conjunction with the bound on $\mathcal{P}^{(0,1)}$ established above, implies that $\min_{\alpha, \beta, m_1, m_2} \mathcal{P}^{(m_1,m_2)}(\alpha, \beta) \simeq -0.101727$.

We now turn to the general case and minimize the integrated probability current $P_{\Psi}$, Eq.~\eqref{P_general-2}, subject to the normalization constraint on $c_m$, Eq.~\eqref{norm-3}. This problem is equivalent to unconstrained minimization of a real-valued functional $I[c_m] = \sum_{m, n = 0}^{\infty} c^*_n K_{nm} c_m - \lambda \sum_{n = 0}^{\infty} c^*_n c_n$, where $\lambda$ is a Lagrange multiplier. The corresponding Euler-Lagrange equation reads
\begin{equation}
	\sum_{n=0}^{\infty} K_{mn} c_n = \lambda c_m \,.
\label{eigenproblem}
\end{equation}
Note that both the matrix $K_{nm}$, defined by Eq.~\eqref{K_def}, and its eigenvalue spectrum $\{ \lambda \}$ depend parametrically on $\alpha$ and $\beta$. Then, the infimum of $P_{\Psi}$ is given by that of the eigenvalue spectrum, i.e.
\begin{equation}
	\mathcal{P}(\alpha,\beta) \equiv \inf_{\Psi} P_{\Psi} = \inf \{ \lambda \} \,.
\label{P_inf_def}
\end{equation} 

In the limit $\alpha \to 0$, which corresponds to $R \to \infty$ and/or $T \to 0$, we recover the Bracken-Melloy bound:
\begin{equation}
	\lim_{\alpha \to 0} \mathcal{P}(\alpha,\beta) \to -c_{\text{line}} \,.
\end{equation}
This can be readily seen by defining $u = m \sqrt{\alpha}$ and $f(u) = c_m / \alpha^{1/4}$, and, for $\alpha \to 0$ and $\beta$ fixed, rewriting Eq.~\eqref{eigenproblem} as
\begin{equation}
	\frac{1}{\pi} \int_0^{\infty} dv \, (u + v) \sinc \left( u^2 - v^2 \right) f(v) = \lambda f(u) \,.
\end{equation}
The last equation is the integral eigenvalue problem originally formulated by Bracken and Melloy \cite{BM94Probability}, and the infimum of its eigenvalue spectrum is $-c_{\text{line}}$.

In general, for arbitrary $\alpha$ and $\beta$, we compute $\mathcal{P}(\alpha,\beta)$ numerically. To this end, we truncate the sum in Eq.~\eqref{eigenproblem} at a large value $n = N$ (of the order of 1000--10000), compute the spectrum $\left\{ \lambda^{(N)} \right\}$ of the corresponding finite-dimensional problem, $\sum_{n=0}^N K_{mn} c_n = \lambda^{(N)} c_m$, and find its minimum $\lambda^{(N)}_{\min} = \min \left\{ \lambda^{(N)} \right\}$. We repeat this calculation for a sequence of increasing $N$, and extrapolate $\lambda^{(N)}_{\min}$ to $N \to \infty$. This procedure yields a numerical estimate for $\mathcal{P}(\alpha,\beta) = \lim_{N \to \infty} \lambda^{(N)}_{\min}$. Further details on the numerical evaluation of $\mathcal{P}$ are provided in Appendix~\ref{appendix}.

\begin{figure}[h]
	\centering
	\includegraphics[width=0.45\textwidth]{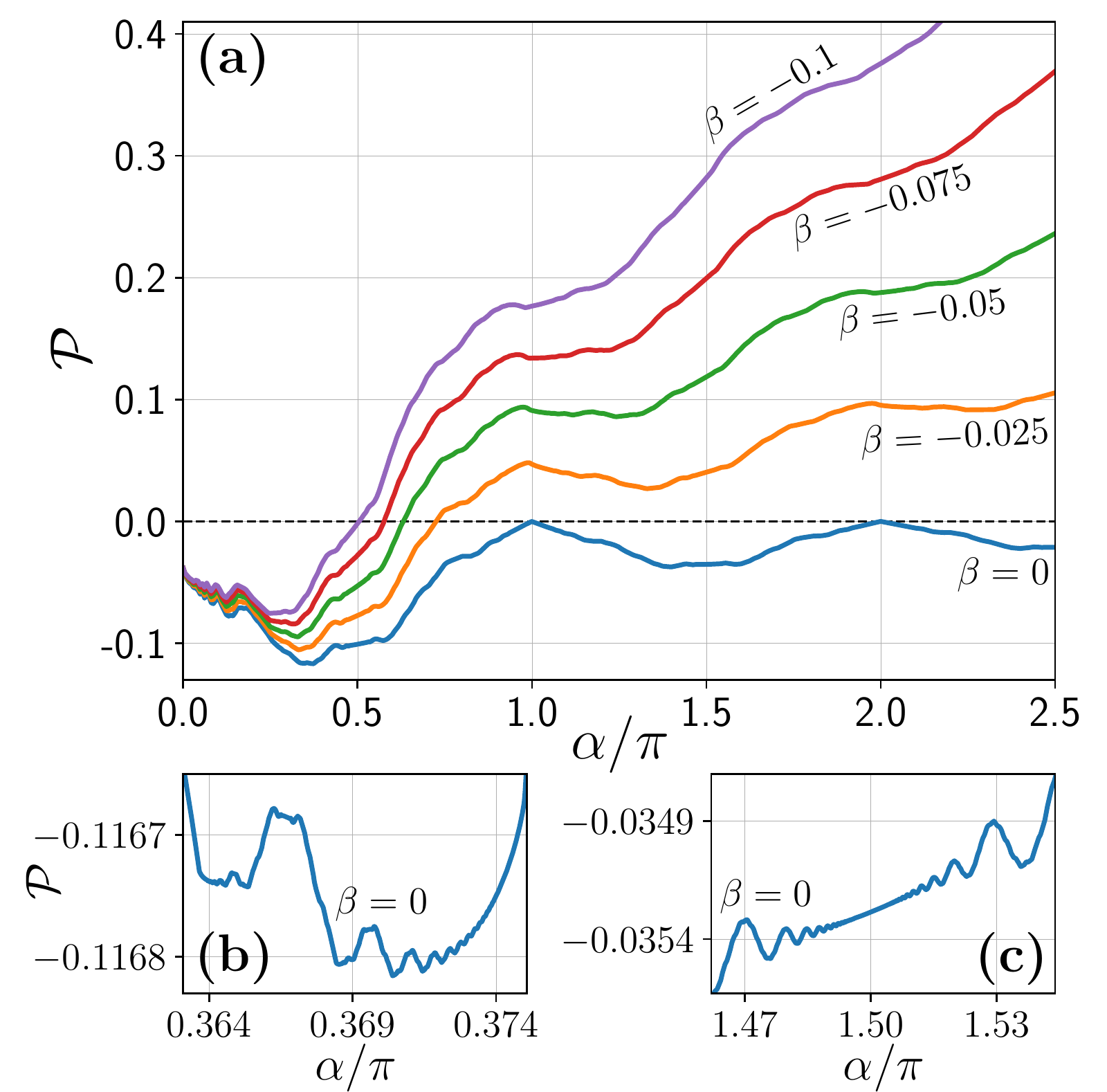}
	\caption{Infimum of the time-integrated probability current, Eq.~\eqref{P_inf_def}, as a function of $\alpha$. (a) $\mathcal{P}(\alpha,\beta)$ for five different values of $\beta$. (b,c) Zoom-ins into $\mathcal{P}(\alpha,0)$.}
	\label{fig2}
\end{figure}

Figure~\ref{fig2}(a) shows the dependence of $\mathcal{P}$ on $\alpha$ for the same five values of $\beta$ as in Fig.~\ref{fig1}. We see that $\mathcal{P}(\alpha,\beta)$ decreases as $\beta$ approaches $0$. It is easy to show that $\mathcal{P}(\alpha,0) = 0$ if $\alpha$ is an integer multiple of $\pi$. For all other values of $\alpha$, the value of $\mathcal{P}(\alpha,0)$ appears to be negative.

Figure~\ref{fig2}(b) shows the curve $\mathcal{P}(\alpha,0)$ in a small interval around $\alpha/\pi \simeq 0.3703965$, where $\mathcal{P}$ attains its smallest value. A careful numerical investigation yields the following estimate for the infimum of the integrated probability current:
\begin{equation}
	 \inf_{\alpha, \beta} \mathcal{P} = -c_{\text{ring}} \,, \qquad c_{\text{ring}} \simeq 0.116816 \,.
\label{main_result_1}
\end{equation}
It is interesting to note that $c_{\text{ring}}$ is more than three times larger than the Bracken-Melloy constant, $c_{\text{line}}$.

Figure~\ref{fig2}(c) is another blow-up of the curve $\mathcal{P}(\alpha,0)$. It illustrates the fact, also evident in Fig.~\ref{fig2}(b), that the dependence of $\mathcal{P}$ on $\alpha$ has an intricate structure on very small scale, as well as some degree of self-similarity. In fact, it might be the case that this dependence has a fractal nature.

We now return to the eigenproblem defined by Eq.~\eqref{eigenproblem} and use it to find a numerical approximation to the backflow-maximizing state. More concretely, we set $\alpha/\pi = 0.3703965$ and $\beta = 0$ (corresponding to $\mathcal{P} \simeq -c_{\text{ring}}$), truncate the sum in Eq.~\eqref{eigenproblem} at $N = 2000$, and compute the eigenvector $(c_0, c_1, \ldots, c_N)$. The sought approximation to the backflow-maximizing state, at $t=0$, is given by $\Psi = \sum_{m=0}^N c_m \psi_m$ [cf.~Eq.~\eqref{wave_function}].

\begin{figure}[h]
	\centering
	\includegraphics[width=0.45\textwidth]{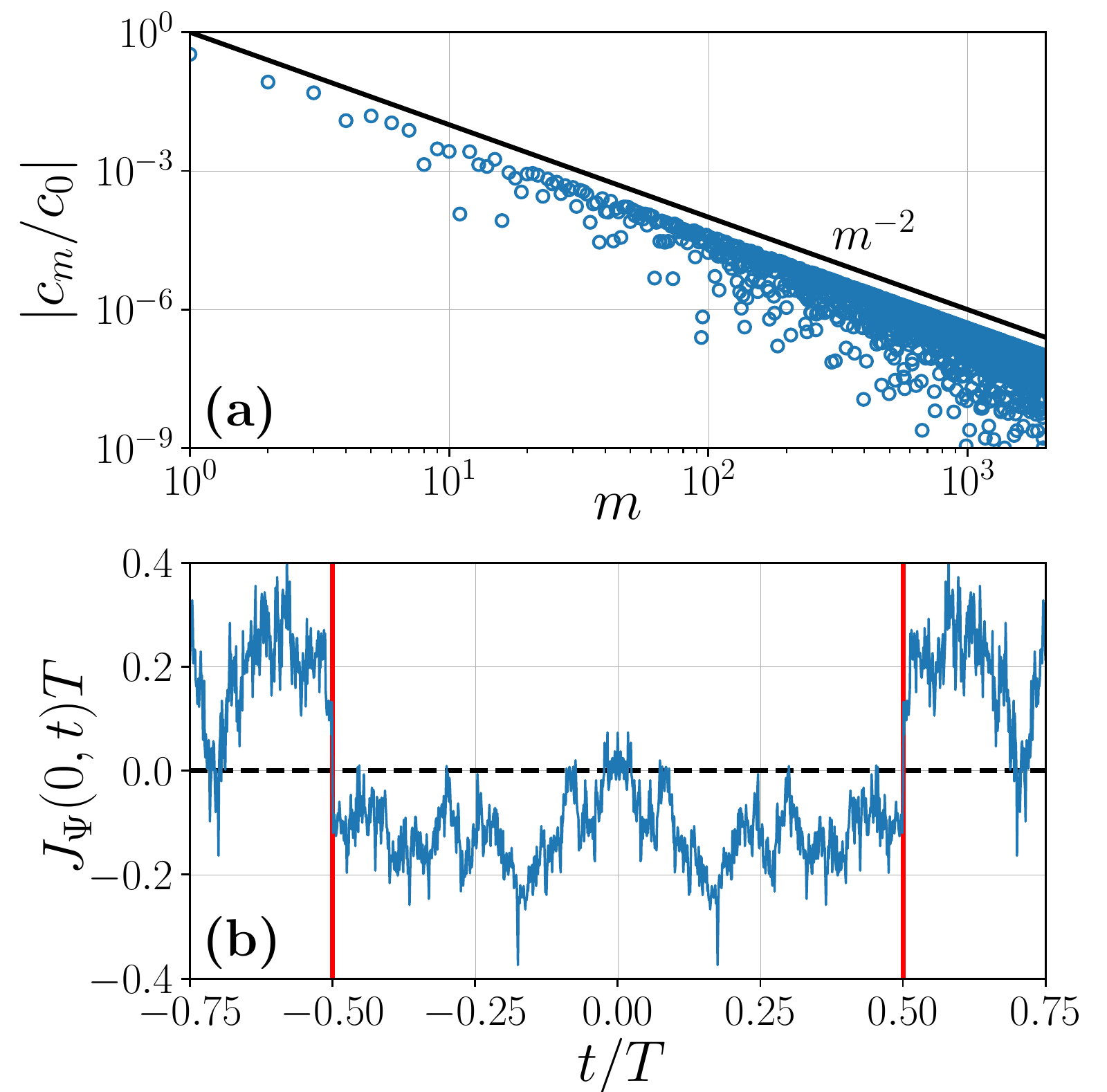}
	\caption{Characteristics of the backflow-maximizing state. (a) Blue circles show the magnitude of the expansion coefficients $c_m$, for $m \ge 1$, in the units of $|c_0|$. The black solid line represents the curve $|c_m/c_0| = m^{-2}$. (b) Probability current $J_{\Psi}(0,t)$ (in units of $1/T$) as a function of time $t$. The interval $-T/2 < t < T/2$ is identified by two red vertical lines.}
	\label{fig3}
\end{figure}

Figure~\ref{fig3}(a) shows, on the log-log scale, the dependence of $|c_m|$ on $m$ for the backflow-maximizing state $\Psi$. We clearly see that
\begin{equation}
	|c_m| < \frac{|c_0|}{m^2} \qquad \forall m \ge 1 \,.
\end{equation}
This inequality ensures that $\Psi$ has a finite mean energy $\langle E \rangle = \sum_m |c_m|^2 E_m$ with $E_m \sim m^2$, as given by Eq.~\eqref{eig}. More precisely, we find
\begin{equation}
	\frac{\langle E \rangle T}{\hbar} \simeq 0.3855 \,.
\end{equation}
This result is in stark contrast to the fact that mean energy of the state maximizing probability backflow on a line is infinite.

We also compute the time-dependent probability current for the backflow-maximizing state $\Psi$. We do this by numerically evaluating the double sum in Eq.~\eqref{prob_current-2} for $\alpha/\pi = 0.3703965$, $\beta = 0$, and $\theta = 0$. Figure~\ref{fig3}(b) shows $J_{\Psi}(0,t)$ (in units of $1/T$) as a function of $t/T$. The integral of $J_{\Psi}(0,t)$ over the time interval $-T/2 < t < T/2$, identified in the figure by two red vertical lines, gives a value close to $-c_{\text{ring}}$. It is interesting to observe that, unlike in the problem of QB on a line, $J_{\Psi}(0,t)$ displays an erratic dependence on time and fails to be everywhere negative on the interval $-T/2 < t < T/2$.

\section{Conclusion}

In conclusion, we have shown that the backflow effect is more pronounced and more amenable to experimental investigation when considered for a particle moving in a circular ring rather than along a straight line. In particular, the integrated backflow current in the ring scenario can be as high as $c_{\text{ring}} \simeq 0.116816$, which is more than three times larger than the corresponding bound in the case of a line, $c_{\text{line}} \simeq 0.0384517$. Also, in the ring case, the energy and spatial extent of the backflow-maximizing state are finite; this gives a significant advantage over the line case in which both of these quantities diverge. Moreover, in the ring case, even very simple states can generate substantial backflow: e.g., a superposition of the ground and first-excited states can yield backflow as high as $87\%$ of the overall bound, $c_{\text{ring}}$. These definite advantages of the particle-in-a-ring system open a new possibility for the first experimental observation of the QB effect.

\begin{acknowledgements}
	The author would like to thank Maximilien Barbier and Remy Dubertrand for helpful discussions.
\end{acknowledgements}

\appendix

\section{Details of the numerical evaluation of~$P(\alpha, \beta)$}
\label{appendix}

Here we give a detailed description of the computational procedure used to evaluate $\mathcal{P}(\alpha, \beta)$. In order to make the description concrete, we take $\alpha / \pi = 0.3703965$ and $\beta = 0$. (These parameter values correspond to $\mathcal{P} \simeq c_{\text{ring}}$.) All numerical calculations were performed in Python and utilized the {\sc numpy} package.

First, using the function {\sc numpy.linalg.eigh}, we compute eigenvalues $\lambda^{(N)}$ of the $N$-dimensional system $\sum_{n=0}^N K_{mn} c_n = \lambda^{(N)} c_m$, and select the smallest eigenvalue $\lambda_{\min}^{(N)}$. We perform this computation for 15 different values of $N$, ranging between 800 and 10000; the results are as follows:
\begin{align*}
	\lambda_{\min}^{(800)} &= -0.11681560946083251 \,, \\
	\lambda_{\min}^{(1000)} &= -0.11681562375295221 \,, \\
	\lambda_{\min}^{(1200)} &= -0.11681563170026898 \,, \\
	\lambda_{\min}^{(1400)} &= -0.11681563657782222 \,, \\
	\lambda_{\min}^{(1600)} &= -0.11681563974451246 \,, \\
	\lambda_{\min}^{(1800)} &= -0.11681564184588990 \,, \\
	\lambda_{\min}^{(2000)} &= -0.11681564340085021 \,, \\
	\lambda_{\min}^{(2200)} &= -0.11681564437173106 \,, \\
	\lambda_{\min}^{(2400)} &= -0.11681564524093137 \,, \\
	\lambda_{\min}^{(3000)} &= -0.11681564684342790 \,, \\
	\lambda_{\min}^{(4000)} &= -0.11681564811514884 \,, \\
	\lambda_{\min}^{(5000)} &= -0.11681564868561355 \,, \\
	\lambda_{\min}^{(6000)} &= -0.11681564900305073 \,, \\
	\lambda_{\min}^{(8000)} &= -0.11681564932805089 \,, \\
	\lambda_{\min}^{(10000)} &= -0.11681564947322964 \,.
\end{align*}

Then, using the function {\sc numpy.polyfit}, $\lambda_{\min}^{(N)}$ is fitted by the following quadratic polynomial in $1/N$: 
\begin{equation}
	\lambda_{\min}^{(N)} \simeq a_0 + \frac{a_1}{N} + \frac{a_2}{N^2}
\label{quad_poly}
\end{equation}
with
\begin{align*}
	a_0 &= -0.11681564972831678 \,, \\
	a_1 &= -5.3630711822449864 \times 10^{-8} \,, \\
	a_2 &= 0.02587490326775755 \,.
\end{align*}
\begin{figure}[h]
	\centering
	\includegraphics[width=0.45\textwidth]{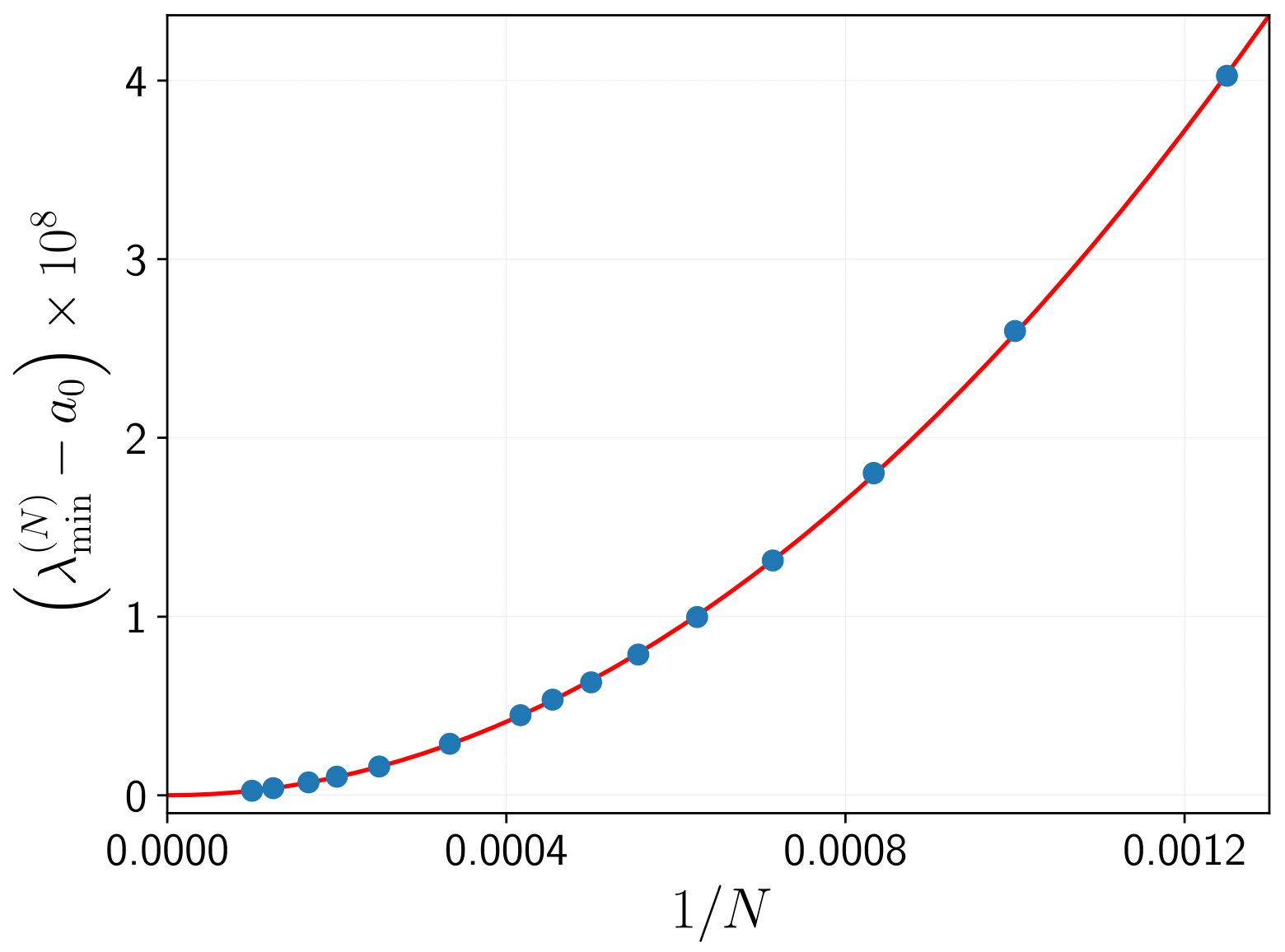}
	\caption{The fit of $\lambda_{\min}^{(N)}$ (blue circles) by the quadratic polynomial in $1/N$ defined by Eq.~\eqref{quad_poly}  (red curve).}
	\label{fig4}
\end{figure}
The fit, shown in Fig.~\ref{fig4}, is very accurate: the corresponding residual (i.e. the sum of the squares of the fit errors) is approximately equal to $7.3 \times 10^{-20}$. This allows us to approximate the sought value of $\mathcal{P} \equiv \lim_{N \to \infty} \lambda_{\min}^{(N)}$ by~$a_0$.

%%%%%%%%%%%%%%%%%%%%%%%%%%%%%%%%%%%%%%%%%%%%%%%%%%%%%%%%%%%%%%%%%%%%%%
%


\begin{thebibliography}{31}%
\makeatletter
\providecommand \@ifxundefined [1]{%
 \@ifx{#1\undefined}
}%
\providecommand \@ifnum [1]{%
 \ifnum #1\expandafter \@firstoftwo
 \else \expandafter \@secondoftwo
 \fi
}%
\providecommand \@ifx [1]{%
 \ifx #1\expandafter \@firstoftwo
 \else \expandafter \@secondoftwo
 \fi
}%
\providecommand \natexlab [1]{#1}%
\providecommand \enquote  [1]{``#1''}%
\providecommand \bibnamefont  [1]{#1}%
\providecommand \bibfnamefont [1]{#1}%
\providecommand \citenamefont [1]{#1}%
\providecommand \href@noop [0]{\@secondoftwo}%
\providecommand \href [0]{\begingroup \@sanitize@url \@href}%
\providecommand \@href[1]{\@@startlink{#1}\@@href}%
\providecommand \@@href[1]{\endgroup#1\@@endlink}%
\providecommand \@sanitize@url [0]{\catcode `\\12\catcode `\$12\catcode
  `\&12\catcode `\#12\catcode `\^12\catcode `\_12\catcode `\%12\relax}%
\providecommand \@@startlink[1]{}%
\providecommand \@@endlink[0]{}%
\providecommand \url  [0]{\begingroup\@sanitize@url \@url }%
\providecommand \@url [1]{\endgroup\@href {#1}{\urlprefix }}%
\providecommand \urlprefix  [0]{URL }%
\providecommand \Eprint [0]{\href }%
\providecommand \doibase [0]{http://dx.doi.org/}%
\providecommand \selectlanguage [0]{\@gobble}%
\providecommand \bibinfo  [0]{\@secondoftwo}%
\providecommand \bibfield  [0]{\@secondoftwo}%
\providecommand \translation [1]{[#1]}%
\providecommand \BibitemOpen [0]{}%
\providecommand \bibitemStop [0]{}%
\providecommand \bibitemNoStop [0]{.\EOS\space}%
\providecommand \EOS [0]{\spacefactor3000\relax}%
\providecommand \BibitemShut  [1]{\csname bibitem#1\endcsname}%
\let\auto@bib@innerbib\@empty
%</preamble>
\bibitem [{\citenamefont {Allcock}(1969)}]{All69time-c}%
  \BibitemOpen
  \bibfield  {author} {\bibinfo {author} {\bibfnamefont {G.~R.}\ \bibnamefont
  {Allcock}},\ }\bibfield  {title} {\enquote {\bibinfo {title} {{The time of
  arrival in quantum mechanics III. The measurement ensemble}},}\ }\href
  {\doibase 10.1016/0003-4916(69)90253-X} {\bibfield  {journal} {\bibinfo
  {journal} {Ann. Phys. (N. Y).}\ }\textbf {\bibinfo {volume} {53}},\ \bibinfo
  {pages} {311} (\bibinfo {year} {1969})}\BibitemShut {NoStop}%
\bibitem [{\citenamefont {Kijowski}(1974)}]{Kij74time}%
  \BibitemOpen
  \bibfield  {author} {\bibinfo {author} {\bibfnamefont {J.}~\bibnamefont
  {Kijowski}},\ }\bibfield  {title} {\enquote {\bibinfo {title} {{On the time
  operator in quantum mechanics and the Heisenberg uncertainty relation for
  energy and time}},}\ }\href {\doibase 10.1016/S0034-4877(74)80004-2}
  {\bibfield  {journal} {\bibinfo  {journal} {Rep. Math. Phys.}\ }\textbf
  {\bibinfo {volume} {6}},\ \bibinfo {pages} {361} (\bibinfo {year}
  {1974})}\BibitemShut {NoStop}%
\bibitem [{\citenamefont {Bracken}\ and\ \citenamefont
  {Melloy}(1994)}]{BM94Probability}%
  \BibitemOpen
  \bibfield  {author} {\bibinfo {author} {\bibfnamefont {A.~J.}\ \bibnamefont
  {Bracken}}\ and\ \bibinfo {author} {\bibfnamefont {G.~F.}\ \bibnamefont
  {Melloy}},\ }\bibfield  {title} {\enquote {\bibinfo {title} {{Probability
  backflow and a new dimensionless quantum number}},}\ }\href {\doibase
  10.1088/0305-4470/27/6/040} {\bibfield  {journal} {\bibinfo  {journal} {J.
  Phys. A: Math. Gen.}\ }\textbf {\bibinfo {volume} {27}},\ \bibinfo {pages}
  {2197} (\bibinfo {year} {1994})}\BibitemShut {NoStop}%
\bibitem [{Note1()}]{Note1}%
  \BibitemOpen
  \bibinfo {note} {See Ref.~\cite {YH13introduction} for a short introduction
  to the quantum backflow effect.}\BibitemShut {Stop}%
\bibitem [{\citenamefont {Eveson}\ \emph {et~al.}(2005)\citenamefont {Eveson},
  \citenamefont {Fewster},\ and\ \citenamefont {Verch}}]{EFV05Quantum}%
  \BibitemOpen
  \bibfield  {author} {\bibinfo {author} {\bibfnamefont {S.~P.}\ \bibnamefont
  {Eveson}}, \bibinfo {author} {\bibfnamefont {C.~J.}\ \bibnamefont {Fewster}},
  \ and\ \bibinfo {author} {\bibfnamefont {R.}~\bibnamefont {Verch}},\
  }\bibfield  {title} {\enquote {\bibinfo {title} {{Quantum Inequalities in
  Quantum Mechanics}},}\ }\href {\doibase 10.1007/s00023-005-0197-9} {\bibfield
   {journal} {\bibinfo  {journal} {Ann. Henri Poincar\'e}\ }\textbf {\bibinfo
  {volume} {6}},\ \bibinfo {pages} {1} (\bibinfo {year} {2005})}\BibitemShut
  {NoStop}%
\bibitem [{\citenamefont {Penz}\ \emph {et~al.}(2006)\citenamefont {Penz},
  \citenamefont {Gr{\"{u}}bl}, \citenamefont {Kreidl},\ and\ \citenamefont
  {Wagner}}]{PGKW06new}%
  \BibitemOpen
  \bibfield  {author} {\bibinfo {author} {\bibfnamefont {M.}~\bibnamefont
  {Penz}}, \bibinfo {author} {\bibfnamefont {G.}~\bibnamefont {Gr{\"{u}}bl}},
  \bibinfo {author} {\bibfnamefont {S.}~\bibnamefont {Kreidl}}, \ and\ \bibinfo
  {author} {\bibfnamefont {P.}~\bibnamefont {Wagner}},\ }\bibfield  {title}
  {\enquote {\bibinfo {title} {{A new approach to quantum backflow}},}\ }\href
  {\doibase 10.1088/0305-4470/39/2/012} {\bibfield  {journal} {\bibinfo
  {journal} {J. Phys. A: Math. Gen.}\ }\textbf {\bibinfo {volume} {39}},\
  \bibinfo {pages} {423} (\bibinfo {year} {2006})}\BibitemShut {NoStop}%
\bibitem [{\citenamefont {Melloy}\ and\ \citenamefont
  {Bracken}(1998{\natexlab{a}})}]{MB98velocity}%
  \BibitemOpen
  \bibfield  {author} {\bibinfo {author} {\bibfnamefont {G.~F.}\ \bibnamefont
  {Melloy}}\ and\ \bibinfo {author} {\bibfnamefont {A.~J.}\ \bibnamefont
  {Bracken}},\ }\bibfield  {title} {\enquote {\bibinfo {title} {{The velocity
  of probability transport in quantum mechanics}},}\ }\href {\doibase
  10.1002/(SICI)1521-3889(199812)7:7/8<726::AID-ANDP726>3.0.CO;2-P} {\bibfield
  {journal} {\bibinfo  {journal} {Ann. Phys. (Leipzig)}\ }\textbf {\bibinfo
  {volume} {7}},\ \bibinfo {pages} {726} (\bibinfo {year}
  {1998}{\natexlab{a}})}\BibitemShut {NoStop}%
\bibitem [{\citenamefont {Muga}\ \emph {et~al.}(1999)\citenamefont {Muga},
  \citenamefont {Palao},\ and\ \citenamefont {Leavens}}]{MPL99Arrival}%
  \BibitemOpen
  \bibfield  {author} {\bibinfo {author} {\bibfnamefont {J.~G.}\ \bibnamefont
  {Muga}}, \bibinfo {author} {\bibfnamefont {J.~P.}\ \bibnamefont {Palao}}, \
  and\ \bibinfo {author} {\bibfnamefont {C.~R.}\ \bibnamefont {Leavens}},\
  }\bibfield  {title} {\enquote {\bibinfo {title} {{Arrival time distributions
  and perfect absorption in classical and quantum mechanics}},}\ }\href
  {\doibase 10.1016/S0375-9601(99)00020-1} {\bibfield  {journal} {\bibinfo
  {journal} {Phys. Lett. A}\ }\textbf {\bibinfo {volume} {253}},\ \bibinfo
  {pages} {21} (\bibinfo {year} {1999})}\BibitemShut {NoStop}%
\bibitem [{\citenamefont {Muga}\ and\ \citenamefont
  {Leavens}(2000)}]{ML00Arrival}%
  \BibitemOpen
  \bibfield  {author} {\bibinfo {author} {\bibfnamefont {J.~G.}\ \bibnamefont
  {Muga}}\ and\ \bibinfo {author} {\bibfnamefont {C.~R.}\ \bibnamefont
  {Leavens}},\ }\bibfield  {title} {\enquote {\bibinfo {title} {{Arrival time
  in quantum mechanics}},}\ }\href {\doibase 10.1016/S0370-1573(00)00047-8}
  {\bibfield  {journal} {\bibinfo  {journal} {Phys. Rep.}\ }\textbf {\bibinfo
  {volume} {338}},\ \bibinfo {pages} {353} (\bibinfo {year}
  {2000})}\BibitemShut {NoStop}%
\bibitem [{\citenamefont {Damborenea}\ \emph {et~al.}(2002)\citenamefont
  {Damborenea}, \citenamefont {Egusquiza}, \citenamefont {Hegerfeldt},\ and\
  \citenamefont {Muga}}]{DEHM02Measurement}%
  \BibitemOpen
  \bibfield  {author} {\bibinfo {author} {\bibfnamefont {J.~A.}\ \bibnamefont
  {Damborenea}}, \bibinfo {author} {\bibfnamefont {I.~L.}\ \bibnamefont
  {Egusquiza}}, \bibinfo {author} {\bibfnamefont {G.~C.}\ \bibnamefont
  {Hegerfeldt}}, \ and\ \bibinfo {author} {\bibfnamefont {J.~G.}\ \bibnamefont
  {Muga}},\ }\bibfield  {title} {\enquote {\bibinfo {title} {{Measurement-based
  approach to quantum arrival times}},}\ }\href {\doibase
  10.1103/PhysRevA.66.052104} {\bibfield  {journal} {\bibinfo  {journal} {Phys.
  Rev. A}\ }\textbf {\bibinfo {volume} {66}},\ \bibinfo {pages} {052104}
  (\bibinfo {year} {2002})}\BibitemShut {NoStop}%
\bibitem [{\citenamefont {Halliwell}\ \emph {et~al.}(2019)\citenamefont
  {Halliwell}, \citenamefont {Beck}, \citenamefont {Lee},\ and\ \citenamefont
  {O'Brien}}]{HBLO19Quasiprobability}%
  \BibitemOpen
  \bibfield  {author} {\bibinfo {author} {\bibfnamefont {J.~J.}\ \bibnamefont
  {Halliwell}}, \bibinfo {author} {\bibfnamefont {H.}~\bibnamefont {Beck}},
  \bibinfo {author} {\bibfnamefont {B.~K.~B.}\ \bibnamefont {Lee}}, \ and\
  \bibinfo {author} {\bibfnamefont {S.}~\bibnamefont {O'Brien}},\ }\bibfield
  {title} {\enquote {\bibinfo {title} {{Quasiprobability for the arrival-time
  problem with links to backflow and the Leggett-Garg inequalities}},}\ }\href
  {\doibase 10.1103/PhysRevA.99.012124} {\bibfield  {journal} {\bibinfo
  {journal} {Phys. Rev. A}\ }\textbf {\bibinfo {volume} {99}},\ \bibinfo
  {pages} {012124} (\bibinfo {year} {2019})}\BibitemShut {NoStop}%
\bibitem [{\citenamefont {Berry}(2010)}]{Ber10Quantum}%
  \BibitemOpen
  \bibfield  {author} {\bibinfo {author} {\bibfnamefont {M.~V.}\ \bibnamefont
  {Berry}},\ }\bibfield  {title} {\enquote {\bibinfo {title} {{Quantum
  backflow, negative kinetic energy, and optical retro-propagation}},}\ }\href
  {\doibase 10.1088/1751-8113/43/41/415302} {\bibfield  {journal} {\bibinfo
  {journal} {J. Phys. A: Math. Theor.}\ }\textbf {\bibinfo {volume} {43}},\
  \bibinfo {pages} {415302} (\bibinfo {year} {2010})}\BibitemShut {NoStop}%
\bibitem [{\citenamefont {Bostelmann}\ \emph {et~al.}(2017)\citenamefont
  {Bostelmann}, \citenamefont {Cadamuro},\ and\ \citenamefont
  {Lechner}}]{BCL17Quantum}%
  \BibitemOpen
  \bibfield  {author} {\bibinfo {author} {\bibfnamefont {H.}~\bibnamefont
  {Bostelmann}}, \bibinfo {author} {\bibfnamefont {D.}~\bibnamefont
  {Cadamuro}}, \ and\ \bibinfo {author} {\bibfnamefont {G.}~\bibnamefont
  {Lechner}},\ }\bibfield  {title} {\enquote {\bibinfo {title} {{Quantum
  backflow and scattering}},}\ }\href {\doibase 10.1103/PhysRevA.96.012112}
  {\bibfield  {journal} {\bibinfo  {journal} {Phys. Rev. A}\ }\textbf {\bibinfo
  {volume} {96}},\ \bibinfo {pages} {012112} (\bibinfo {year}
  {2017})}\BibitemShut {NoStop}%
\bibitem [{\citenamefont {Melloy}\ and\ \citenamefont
  {Bracken}(1998{\natexlab{b}})}]{MB98Probability}%
  \BibitemOpen
  \bibfield  {author} {\bibinfo {author} {\bibfnamefont {G.~F.}\ \bibnamefont
  {Melloy}}\ and\ \bibinfo {author} {\bibfnamefont {A.~J.}\ \bibnamefont
  {Bracken}},\ }\bibfield  {title} {\enquote {\bibinfo {title} {{Probability
  Backflow for a Dirac Particle}},}\ }\href {\doibase 10.1023/A:1018724313788}
  {\bibfield  {journal} {\bibinfo  {journal} {Found. Phys.}\ }\textbf {\bibinfo
  {volume} {28}},\ \bibinfo {pages} {505} (\bibinfo {year}
  {1998}{\natexlab{b}})}\BibitemShut {NoStop}%
\bibitem [{\citenamefont {Su}\ and\ \citenamefont {Chen}(2018)}]{SC18Quantum}%
  \BibitemOpen
  \bibfield  {author} {\bibinfo {author} {\bibfnamefont {H.}~\bibnamefont
  {Su}}\ and\ \bibinfo {author} {\bibfnamefont {J.}~\bibnamefont {Chen}},\
  }\bibfield  {title} {\enquote {\bibinfo {title} {{Quantum backflow in
  solutions to the Dirac equation of the spin-1/2 free particle}},}\ }\href
  {\doibase 10.1142/S0217732318501869} {\bibfield  {journal} {\bibinfo
  {journal} {Mod. Phys. Lett. A}\ }\textbf {\bibinfo {volume} {33}},\ \bibinfo
  {pages} {1850186} (\bibinfo {year} {2018})}\BibitemShut {NoStop}%
\bibitem [{\citenamefont {Ashfaque}\ \emph {et~al.}(2019)\citenamefont
  {Ashfaque}, \citenamefont {Lynch},\ and\ \citenamefont
  {Strange}}]{ALS19Relativistic}%
  \BibitemOpen
  \bibfield  {author} {\bibinfo {author} {\bibfnamefont {J.~M.}\ \bibnamefont
  {Ashfaque}}, \bibinfo {author} {\bibfnamefont {J.}~\bibnamefont {Lynch}}, \
  and\ \bibinfo {author} {\bibfnamefont {P.}~\bibnamefont {Strange}},\
  }\bibfield  {title} {\enquote {\bibinfo {title} {{Relativistic quantum
  backflow}},}\ }\href {\doibase 10.1088/1402-4896/ab265c} {\bibfield
  {journal} {\bibinfo  {journal} {Phys. Scr.}\ }\textbf {\bibinfo {volume}
  {94}},\ \bibinfo {pages} {125107} (\bibinfo {year} {2019})}\BibitemShut
  {NoStop}%
\bibitem [{\citenamefont {Goussev}(2019)}]{Gou19Equivalence}%
  \BibitemOpen
  \bibfield  {author} {\bibinfo {author} {\bibfnamefont {A.}~\bibnamefont
  {Goussev}},\ }\bibfield  {title} {\enquote {\bibinfo {title} {{Equivalence
  between quantum backflow and classically forbidden probability flow in a
  diffraction-in-time problem}},}\ }\href {\doibase 10.1103/PhysRevA.99.043626}
  {\bibfield  {journal} {\bibinfo  {journal} {Phys. Rev. A}\ }\textbf {\bibinfo
  {volume} {99}},\ \bibinfo {pages} {043626} (\bibinfo {year}
  {2019})}\BibitemShut {NoStop}%
\bibitem [{\citenamefont {van Dijk}\ and\ \citenamefont
  {Toyama}(2019)}]{DT19Decay}%
  \BibitemOpen
  \bibfield  {author} {\bibinfo {author} {\bibfnamefont {W.}~\bibnamefont {van
  Dijk}}\ and\ \bibinfo {author} {\bibfnamefont {F.~M.}\ \bibnamefont
  {Toyama}},\ }\bibfield  {title} {\enquote {\bibinfo {title} {{Decay of a
  quasistable quantum system and quantum backflow}},}\ }\href {\doibase
  10.1103/PhysRevA.100.052101} {\bibfield  {journal} {\bibinfo  {journal}
  {Phys. Rev. A}\ }\textbf {\bibinfo {volume} {100}},\ \bibinfo {pages}
  {052101} (\bibinfo {year} {2019})}\BibitemShut {NoStop}%
\bibitem [{\citenamefont {Barbier}(2020)}]{Bar20Quantum}%
  \BibitemOpen
  \bibfield  {author} {\bibinfo {author} {\bibfnamefont {M.}~\bibnamefont
  {Barbier}},\ }\bibfield  {title} {\enquote {\bibinfo {title} {{Quantum
  backflow for many-particle systems}},}\ }\href@noop {} {\bibfield  {journal}
  {\bibinfo  {journal} {arXiv:2005.14685}\ } (\bibinfo {year}
  {2020})}\BibitemShut {NoStop}%
\bibitem [{\citenamefont {Goussev}(2020)}]{Gou20Probability}%
  \BibitemOpen
  \bibfield  {author} {\bibinfo {author} {\bibfnamefont {A.}~\bibnamefont
  {Goussev}},\ }\bibfield  {title} {\enquote {\bibinfo {title} {{Probability
  backflow for correlated quantum states}},}\ }\href {\doibase
  10.1103/PhysRevResearch.2.033206} {\bibfield  {journal} {\bibinfo  {journal}
  {Phys. Rev. Research}\ }\textbf {\bibinfo {volume} {2}},\ \bibinfo {pages}
  {033206} (\bibinfo {year} {2020})}\BibitemShut {NoStop}%
\bibitem [{Note2()}]{Note2}%
  \BibitemOpen
  \bibinfo {note} {An experimental scheme for observing QB in Bose-Einstein
  condensates was proposed in Ref.~\cite {PTMM13Detecting}. Also, there has
  been a recent experimental realization of an optical analog of the QB
  effect \cite {EZB20Observation}.}\BibitemShut {Stop}%
\bibitem [{\citenamefont {Albarelli}\ \emph {et~al.}(2016)\citenamefont
  {Albarelli}, \citenamefont {Guaita},\ and\ \citenamefont
  {Paris}}]{AGP16Quantum}%
  \BibitemOpen
  \bibfield  {author} {\bibinfo {author} {\bibfnamefont {F.}~\bibnamefont
  {Albarelli}}, \bibinfo {author} {\bibfnamefont {T.}~\bibnamefont {Guaita}}, \
  and\ \bibinfo {author} {\bibfnamefont {M.~G.~A.}\ \bibnamefont {Paris}},\
  }\bibfield  {title} {\enquote {\bibinfo {title} {{Quantum backflow effect and
  nonclassicality}},}\ }\href {\doibase 10.1142/S0219749916500325} {\bibfield
  {journal} {\bibinfo  {journal} {Int. J. Quantum Inf.}\ }\textbf {\bibinfo
  {volume} {14}},\ \bibinfo {pages} {1650032} (\bibinfo {year}
  {2016})}\BibitemShut {NoStop}%
\bibitem [{\citenamefont {Yearsley}\ \emph {et~al.}(2012)\citenamefont
  {Yearsley}, \citenamefont {Halliwell}, \citenamefont {Hartshorn},\ and\
  \citenamefont {Whitby}}]{YHHW12Analytical}%
  \BibitemOpen
  \bibfield  {author} {\bibinfo {author} {\bibfnamefont {J.~M.}\ \bibnamefont
  {Yearsley}}, \bibinfo {author} {\bibfnamefont {J.~J.}\ \bibnamefont
  {Halliwell}}, \bibinfo {author} {\bibfnamefont {R.}~\bibnamefont
  {Hartshorn}}, \ and\ \bibinfo {author} {\bibfnamefont {A.}~\bibnamefont
  {Whitby}},\ }\bibfield  {title} {\enquote {\bibinfo {title} {{Analytical
  examples, measurement models, and classical limit of quantum backflow}},}\
  }\href {\doibase 10.1103/PhysRevA.86.042116} {\bibfield  {journal} {\bibinfo
  {journal} {Phys. Rev. A}\ }\textbf {\bibinfo {volume} {86}},\ \bibinfo
  {pages} {042116} (\bibinfo {year} {2012})}\BibitemShut {NoStop}%
\bibitem [{\citenamefont {Halliwell}\ \emph {et~al.}(2013)\citenamefont
  {Halliwell}, \citenamefont {Gillman}, \citenamefont {Lennon}, \citenamefont
  {Patel},\ and\ \citenamefont {Ramirez}}]{HGL+13Quantum}%
  \BibitemOpen
  \bibfield  {author} {\bibinfo {author} {\bibfnamefont {J.~J.}\ \bibnamefont
  {Halliwell}}, \bibinfo {author} {\bibfnamefont {E.}~\bibnamefont {Gillman}},
  \bibinfo {author} {\bibfnamefont {O.}~\bibnamefont {Lennon}}, \bibinfo
  {author} {\bibfnamefont {M.}~\bibnamefont {Patel}}, \ and\ \bibinfo {author}
  {\bibfnamefont {I.}~\bibnamefont {Ramirez}},\ }\bibfield  {title} {\enquote
  {\bibinfo {title} {{Quantum backflow states from eigenstates of the
  regularized current operator}},}\ }\href {\doibase
  10.1088/1751-8113/46/47/475303} {\bibfield  {journal} {\bibinfo  {journal}
  {J. Phys. A: Math. Theor.}\ }\textbf {\bibinfo {volume} {46}},\ \bibinfo
  {pages} {475303} (\bibinfo {year} {2013})}\BibitemShut {NoStop}%
\bibitem [{\citenamefont {Strange}(2012)}]{Str12Large}%
  \BibitemOpen
  \bibfield  {author} {\bibinfo {author} {\bibfnamefont {P.}~\bibnamefont
  {Strange}},\ }\bibfield  {title} {\enquote {\bibinfo {title} {{Large quantum
  probability backflow and the azimuthal angle–angular momentum uncertainty
  relation for an electron in a constant magnetic field}},}\ }\href {\doibase
  10.1088/0143-0807/33/5/1147} {\bibfield  {journal} {\bibinfo  {journal} {Eur.
  J. Phys.}\ }\textbf {\bibinfo {volume} {33}},\ \bibinfo {pages} {1147}
  (\bibinfo {year} {2012})}\BibitemShut {NoStop}%
\bibitem [{\citenamefont {Paccoia}\ \emph {et~al.}(2020)\citenamefont
  {Paccoia}, \citenamefont {Panella},\ and\ \citenamefont
  {Roy}}]{PPR20Angular}%
  \BibitemOpen
  \bibfield  {author} {\bibinfo {author} {\bibfnamefont {V.~D.}\ \bibnamefont
  {Paccoia}}, \bibinfo {author} {\bibfnamefont {O.}~\bibnamefont {Panella}}, \
  and\ \bibinfo {author} {\bibfnamefont {P.}~\bibnamefont {Roy}},\ }\bibfield
  {title} {\enquote {\bibinfo {title} {Angular momentum quantum backflow in the
  noncommutative plane},}\ }\href@noop {} {\bibfield  {journal} {\bibinfo
  {journal} {Phys. Rev. A (in press), arXiv:2011.13644}\ } (\bibinfo {year}
  {2020})}\BibitemShut {NoStop}%
\bibitem [{\citenamefont {Scardicchio}(2002)}]{Sca02Classical}%
  \BibitemOpen
  \bibfield  {author} {\bibinfo {author} {\bibfnamefont {A.}~\bibnamefont
  {Scardicchio}},\ }\bibfield  {title} {\enquote {\bibinfo {title} {{Classical
  and quantum dynamics of a particle constrained on a circle}},}\ }\href
  {\doibase 10.1016/S0375-9601(02)00690-4} {\bibfield  {journal} {\bibinfo
  {journal} {Phys. Lett. A}\ }\textbf {\bibinfo {volume} {300}},\ \bibinfo
  {pages} {7} (\bibinfo {year} {2002})}\BibitemShut {NoStop}%
\bibitem [{\citenamefont {Vugalter}\ \emph {et~al.}(2004)\citenamefont
  {Vugalter}, \citenamefont {Das},\ and\ \citenamefont
  {Sorokin}}]{Vug04carged}%
  \BibitemOpen
  \bibfield  {author} {\bibinfo {author} {\bibfnamefont {G.~A.}\ \bibnamefont
  {Vugalter}}, \bibinfo {author} {\bibfnamefont {A.~K.}\ \bibnamefont {Das}}, \
  and\ \bibinfo {author} {\bibfnamefont {V.~A.}\ \bibnamefont {Sorokin}},\
  }\bibfield  {title} {\enquote {\bibinfo {title} {{A charged particle on a
  ring in a magnetic field: quantum revivals}},}\ }\href {\doibase
  10.1088/0143-0807/25/2/003} {\bibfield  {journal} {\bibinfo  {journal} {Eur.
  J. Phys.}\ }\textbf {\bibinfo {volume} {25}},\ \bibinfo {pages} {157}
  (\bibinfo {year} {2004})}\BibitemShut {NoStop}%
\bibitem [{\citenamefont {Yearsley}\ and\ \citenamefont
  {Halliwell}(2013)}]{YH13introduction}%
  \BibitemOpen
  \bibfield  {author} {\bibinfo {author} {\bibfnamefont {J.~M.}\ \bibnamefont
  {Yearsley}}\ and\ \bibinfo {author} {\bibfnamefont {J.~J.}\ \bibnamefont
  {Halliwell}},\ }\bibfield  {title} {\enquote {\bibinfo {title} {{An
  introduction to the quantum backflow effect}},}\ }\href {\doibase
  10.1088/1742-6596/442/1/012055} {\bibfield  {journal} {\bibinfo  {journal}
  {J. Phys. Conf. Ser.}\ }\textbf {\bibinfo {volume} {442}},\ \bibinfo {pages}
  {012055} (\bibinfo {year} {2013})}\BibitemShut {NoStop}%
\bibitem [{\citenamefont {Palmero}\ \emph {et~al.}(2013)\citenamefont
  {Palmero}, \citenamefont {Torrontegui}, \citenamefont {Muga},\ and\
  \citenamefont {Modugno}}]{PTMM13Detecting}%
  \BibitemOpen
  \bibfield  {author} {\bibinfo {author} {\bibfnamefont {M.}~\bibnamefont
  {Palmero}}, \bibinfo {author} {\bibfnamefont {E.}~\bibnamefont
  {Torrontegui}}, \bibinfo {author} {\bibfnamefont {J.~G.}\ \bibnamefont
  {Muga}}, \ and\ \bibinfo {author} {\bibfnamefont {M.}~\bibnamefont
  {Modugno}},\ }\bibfield  {title} {\enquote {\bibinfo {title} {{Detecting
  quantum backflow by the density of a Bose-Einstein condensate}},}\ }\href
  {\doibase 10.1103/PhysRevA.87.053618} {\bibfield  {journal} {\bibinfo
  {journal} {Phys. Rev. A}\ }\textbf {\bibinfo {volume} {87}},\ \bibinfo
  {pages} {053618} (\bibinfo {year} {2013})}\BibitemShut {NoStop}%
\bibitem [{\citenamefont {Eliezer}\ \emph {et~al.}(2020)\citenamefont
  {Eliezer}, \citenamefont {Zacharias},\ and\ \citenamefont
  {Bahabad}}]{EZB20Observation}%
  \BibitemOpen
  \bibfield  {author} {\bibinfo {author} {\bibfnamefont {Y.}~\bibnamefont
  {Eliezer}}, \bibinfo {author} {\bibfnamefont {T.}~\bibnamefont {Zacharias}},
  \ and\ \bibinfo {author} {\bibfnamefont {A.}~\bibnamefont {Bahabad}},\
  }\bibfield  {title} {\enquote {\bibinfo {title} {{Observation of optical
  backflow}},}\ }\href {\doibase 10.1364/OPTICA.371494} {\bibfield  {journal}
  {\bibinfo  {journal} {Optica}\ }\textbf {\bibinfo {volume} {7}},\ \bibinfo
  {pages} {72} (\bibinfo {year} {2020})}\BibitemShut {NoStop}%
\end{thebibliography}
\end{document}